\documentclass[a4paper]{jpconf}
\usepackage{graphicx}

\newcommand{\eq}[1]{Eq.~(\ref{#1})}
\newcommand{\eqs}[1]{Eqs~(\ref{#1})}
\newcommand{\ud}{\ensuremath{\mathrm{d}}}
\newcommand{\ie}{\textit{i.e.}}
\newcommand{\eg}{\textit{e.g.}}
\newcommand{\etc}{\textit{etc}}

\begin{document}
\title{Nonuniqueness of stationary accretion in the Shakura model}
\author{Edward Malec and Krzysztof Roszkowski}
\address{M. Smoluchowski Institute of Physics,
Jagellonian University, Reymonta 4, 30-059 Krak\'{o}w, Poland}
\ead{malec@th.if.uj.edu.pl, roszkowski@th.if.uj.edu.pl}

\begin{abstract}
We investigate the accretion onto luminous bodies with hard surfaces within the
framework of newtonian theory. The accreting gases are assumed to be polytropes
and their selfgravitation is included. A remarkable feature of the model is
that under proper boundary conditions some parameters of the sonic point are
the same as in the Bondi model and the relation between luminosity and the gas
abundance reduces to an algebraic relation. All that holds under assumptions
of stationarity and spherical symmetric. Assuming data that are required for
the complete specification of the system, one finds that generically for
a given luminosity there exist two solutions with different compact cores.
\end{abstract}

\section{Introduction}
Stationary accretion of spherically symmetric fluids with luminosities close
to the Eddington limit has been investigated since the pioneering work of
Shakura \cite{Shakura}. In \cite{Shakura} the gas pressure and its selfgravity
have been ignored and the analysis has been purely newtonian. Later researchers
included both these aspects \cite{later} and extended the analysis onto
relativistic systems \cite{Thorne}.

This paper deals with the complete newtonian picture (which we will refer to as
the Shakura model), in which pressure and selfgravity are included. It is
assumed that a ball of a polytropic selfgravitating gas accretes stationarily
onto a body with a hard surface; both are spherically symmetric. Our goal is to
study a kind of an \textit{inverse problem}. Let be given the luminosity,
total mass, asymptotic temperature and the equation of state of the gas. Assume
that the gravitational potential has a fixed value at the boundary of a compact
body, but the radius of the body is arbitrary to some extent. One can easily
check that these data are sufficient to determine solutions of the Shakura
model; any additional information would lead to a contradiction. On the other
hand, it is well known that nonlinear problems can exhibit nonuniqueness.
In our concrete example the nonuniqueness would manifest itself in multiple
solutions which have different central compact cores. That issue is interesting
in itself within the framework of astrophysics. The fundamental question is
whether observations can distinguish compact bodies endowed with a hard surface
(say, gravastars \cite{Mazur}) from black holes. Narayan \cite{Narayan}
presented observational arguments in favour of the existence of black holes.
Abramowicz, Klu\'zniak and Lasota \cite{Abramowicz} raised several objections,
pointing out that present accretion disk models are not capable to make
distinction between accretion onto a black hole or a gravastar.

The Shakura model is the simplest selfcontained system that can be interpreted
as a radiating system. Since we work at the newtonian level, the issue of
making distinction between a black hole and a compact body with a hard surface
is not addressed. Our results show, however, that there is an ambiguity.
There can exists at least two systems (two compact stars with a hard surface,
having different masses) to the \textit{inverse problem} formulated above.
This is consistent with the recent general-relativistic analysis
\cite{Kinasiewicz, KMM}, valid in the newtonian limit (with a reservation),
which shows that the mass accretion rate $\dot M$ behaves like $y^2(1-y)$.
Here $y = M_\ast / M$ is the ratio of (roughly) the central mass to
the total mass. Therefore the maximum of $\dot M$ is achieved at $y=2/3$.
The luminosity is equal to the mass accretion rate $\dot M$ multiplied by the
available energy per unit mass, the potential $\phi (R_0)$, at the hard surface
of a compact star. Since the boundary value $\phi (R_0)$ is fixed, this would
mean that there appear two weakly luminous regimes: one rich in fluid with
$m_f \approx M$ and the other with a small amount of fluid, $m_f/M\ll 1$.
That would suggest a nonuniqueness~--- given a luminosity $x$, one would have
two systems with different $y$'s. The situation in the Shakura model is not so
simple, because the luminosity impacts also the accretion rate $\dot M$ and
there emerges a complex functional relation $y=y(x)$. Our analysis shows that
under suitable boundary conditions this relation $y=y(x)$ is well approximated
by an algebraic equation, and then one shows that the nonuniqueness again
emerges in the Shakura model.

\section{Main equations}
We will study a spherically symmetric stationary flow of a fluid onto
a spherical compact body. The areal velocity $U$ of a comoving particle labeled
by coordinates $(r,t)$ is given by $U(r,t) = \partial_t R $, where $t$ is a
comoving (Lagrangian) time. $p$ denotes pressure, $L(R)$ and $L_E$ are the
local and Eddington luminosities, respectively.  The mass $m(R)$ is given by
$\partial_R m(R) = 4 \pi R^2 \varrho $. The mass accretion rate reads
\begin{equation}
\dot M = -4 \pi R^2 \varrho U,
\label{dot}
\end{equation}
where $\varrho $ is the baryonic mass density. Since the fluid is a polytrope,
we have the equation of state $p = K \varrho^\Gamma$ with a constant $\Gamma$,
($1 < \Gamma \le 5/3$). The steadiness of the collapse means that all
characteristics of the fluid are constant at a fixed areal radius
$R$: $\partial_t X|_{R = \mathrm{const}}
 = (\partial_t - (\partial_t R) \partial_R) X = 0$,
where $X = \varrho, U, a^2$. The speed of sound is given by
$a=\sqrt{\partial_\varrho p} $. Strictly saying, a stationary accretion must
lead to to the increase of the central mass. This in turn means that the notion
of the ``steady accretion'' is approximate~--- it demands that the mass
accretion rate is small and the time scale is short enough, so that the
quasilocal mass $m(R)$ does not change significantly. The total mass
$m(R_\infty )$ will be denoted by $M$. It is assumed that the radius $R_\infty$
of the ball of fluid and other boundary data are such that
\begin{equation}
U^2_{\infty} \ll \frac{G m(R_\infty)}{R_\infty}  \ll a^2_\infty .
\label{cond}
\end{equation}
The value of the gravitational potential at the surface of the compact body
is fixed to be $-\phi_0$.

The steadily accreting polytropic fluid is described by a system of nonlinear
integro-differential equations. They consist of the Euler equation
\begin{equation}
U\partial_R U =
-{Gm(R)\over R^2} -{1\over \varrho } \partial_R p + \alpha {L(r) \over R^2},
\label{1}
\end{equation}
the mass conservation
\begin{equation}
\partial_R \dot M =0 ,
\label{2}
\end{equation}
and the energy conservation
\begin{equation}
L_0-L(r) =
\dot M \left( {a^2_\infty \over \Gamma -1} - {a^2 \over \Gamma -1}
-{U^2\over 2} -\phi (R) \right) .
\label{3}
\end{equation}
The constant $\alpha $ is given by $\alpha =\sigma_T /4\pi m_p c$, $L_0$ is the
total luminosity and $\phi (R) $ is the newtonian gravitational potential.
We neglected in \eq{3} the term ${U^2_\infty \over 2}-{M\over R_\infty }$ since
it is much smaller than ${a^2_\infty \over \Gamma -1}$, due to the boundary
conditions that have been displayed above.

The Eddington luminosity is calculated here for the whole system, that is
$L_E=GM/\alpha $, while the total luminosity $L_0$ is equal to the product of
the mass accretion rate and $\phi_0\equiv |\phi (R_0)|$, where $R_0$ is an
areal radius of the boundary of the compact body. $L_0=\dot M \phi_0$
\cite{Shakura}. The quantity $\phi_0$ is fixed, as said above, and $R_0$ in the
constructed configurations is by definition the radius at which the absolute
value of the potential equals to $\phi_0$. $R_0$ can be found only after the
solution of the whole system is constructed. After some algebra (which consists
of differentiating (\ref{2}) with respect to $R$, using the equality
$\partial_R a^2/(\Gamma -1) = a^2\partial_R \ln \varrho =\partial_R p/\varrho$
and combining the obtained equation with (\ref{1})) one arrives at
\begin{equation}
\partial_R\ln L = {\alpha \dot M \over R^2}.
\label{4}
\end{equation}
This can be immediately solved,
\begin{equation}
L = L_0\exp \left( {-L_0\tilde R_0\over L_ER}\right) .
\label{5}
\end{equation}
Here appears a new quantity $\tilde R_0 \equiv GM/|\phi (R_0)|$. $\tilde R_0$
represents a kind of the size measure of the compact body. For test fluids
(when the absolute value of the potential is just the product of the
gravitational constant $G$ by the mass divided by the radius) one has
$\tilde R_0=R_0$, while in the general case $\tilde R > R_0$.

One can prove, using the same arguments that Bondi applied to the accreting
newtonian stars with test gases \cite{Bondi}, that there exists a unique
solution in the case without selfgravitation. We believe that this is just
a matter of a simple (and perhaps tedious) work to prove analytically the
existence of solutions of the Shakura model. In any case, there exist numerical
solutions~\cite{Karkowski06a}.

\section{The mass accretion rate}
Below we assume that the accretion flow possesses the so-called sonic point,
but in our opinion the same conclusions should be valid for flows without sonic
points. The standard terminology~--- critical flows (if there is a sonic point)
and subcritical flows (if the sonic point is absent)~--- will not be used in
this paper. The notions ``critical solutions'' and ``subcritical solutions''
will appear, but in a different meaning, defined later.

The sonic horizon (sonic point) is at a location where $|U| = a$. In the
following we will denote by the asterisk all values referring to the sonic
points, \eg, $a_\ast$, $U_\ast $, \etc. Differentiation with respect to the
areal radius will be denoted as prime $'$. The mass conservation equation yields
\begin{equation}
U'=-U\left( \varrho '/\varrho + 2/R\right) .
\label{u}
\end{equation}
Inserting (\ref{u}) into \eq{1} one arrives at
\begin{equation}
{\varrho '\over \varrho } (a^2-U^2)= {2\over R} \left(U^2-{Gm(R)\over 2R }+
{L(R)\alpha \over 2R} \right).
\label{6}
\end{equation}
Using the definition of the sonic point one discovers that its three
characteristics, $a_\ast$, $U_\ast$ and $M_\ast / R_\ast$ must satisfy following
relations
\begin{equation}
a_\ast^2 =
U_\ast^2 =
{G M_\ast \over 2R_\ast }\left( 1- {L_\ast \alpha \over GM_\ast }\right) =
{G M_\ast \over 2R_\ast }\left( 1- {L_\ast M \over L_EM_\ast }\right).
\label{7}
\end{equation}
\eq{7} implies that if there exists a sonic point then
\begin{equation}
{L_\ast M\over L_EM_\ast }<1.
\label{cf}
\end{equation}
The velocity $U$ and the mass density $\varrho $ can be expressed as follows
\begin{equation}
U =
U_\ast \frac{R^2_\ast}{R^2} \left( {a^2_\ast \over a^2} \right)^{1/(\Gamma -1)},
\label{U}
\end{equation}
and
\begin{equation}
\varrho = \varrho_{\infty } \left( a/a_\infty \right)^{2/(\Gamma - 1)}.
\label{rho}
\end{equation}
Here $U_\ast$ is the negative square root and the constant $\varrho_\infty$ is
the asymptotic mass density of a collapsing fluid. Two of the four classes of
stationary solutions that cross at the sonic point can be assigned physical
meaning of an accretion or a wind.

Let us introduce auxiliary variables
\begin{eqnarray}
&& x \equiv {L_0\over L_E}, \qquad
y\equiv {M_\ast \over M }, \qquad
\gamma \equiv {\tilde R_0\over R_\ast }, \nonumber\\
&& \Delta_* \equiv
L_0-L_\ast -2\dot M a^2_\ast {L_\ast \over yL_E- L_\ast },\nonumber\\
&& \Psi_\ast \equiv -\phi_\ast -{GM_\ast \over R_\ast}.
\label{def}
\end{eqnarray}
The necessary condition (\ref{cf}) for the existence of a sonic point can be now
formulated as
\begin{equation}
x\exp \left( -x\gamma \right) < y.
\label{cf2}
\end{equation}
A straightforward calculation yields the following equation
\begin{equation}
a^2_\infty -{5 - 3 \Gamma \over 2}a^2_\ast =\left( \Gamma -1\right)
\left( {\Delta_\ast \over \dot M} -\Psi_\ast \right) .
\label{theorem2a}
\end{equation}
We shall assume that $\gamma <1 $ and $x\gamma \ll 1$.

\eq{7} implies that
\begin{equation}
a^2_\ast ={GM_\ast \over 2R_\ast }\left( 1- { x\over y }
\exp\left( -x\gamma \right)\right) \approx
{GM_\ast \over 2R_\ast }\left( 1- { x\over y } \right) .
\label{y}
\end{equation}
One obtains from (\ref{def}) (keeping only the first term in the expansion of
the luminosity function~$L_\ast $)
\begin{eqnarray}
L_0-L_\ast & = &xL_0{GM\over |R_\ast \phi (R_0)|} =
{x\over y} {\dot M GM_\ast \over R_\ast }=
\nonumber \\
& = &2\dot M a^2_\ast {x\over y-x}.
\end{eqnarray}
In the same approximation the third term in the expression for  $\Delta_\ast$
reads $-2\dot M a^2_\ast {x\over y-x}$; therefore
$\Delta_\ast /\dot M \approx 0$. One can show, using a reasoning similar to
that provided in the next section, that
$\Psi_\ast =4G\pi \int_{R_\ast }^{R_\infty}\ud r\, r \varrho $ is much smaller
than $a^2_\ast$.

Thence the sonic point formula (\ref{theorem2a}) simplifies to
\begin{equation}
{a^2_\ast \over a^2_\infty } = {2\over 5 - 3 \Gamma }.
\label{theorem2}
\end{equation}
That is one of the main results of this presentation. The reader can be
surprised by the fact that we obtained the same result as in the Bondi model,
in which the luminosity and selfgravitation are absent \cite{Bondi}. The
explanation for that lies in our boundary conditions (\ref{cond}). Relaxing them
(for instance assuming a nonzero asymptotic falloff velocity $U$) would
invalidate formula (\ref{theorem2}) and also the remaining results of this work.
We would also like to warn the reader that it is assumed in the above derivation
that $\Gamma $ is somewhat smaller than 5/3.

\section{Luminosity}
In this section we return to the question stated at the beginning. Namely, let
be given the total mass $M$, the size $R_\infty $, the luminosity $L_0$ and the
asymptotic temperature $T$ (or equivalently, the asymptotic speed of sound
$a^2_\infty $). We show that in the generic case there exist two accreting
systems.

The rate of the mass accretion $\dot M$ (see \eq{dot}) can be formulated in
terms of the characteristics of the sonic point. After some algebra one derives
from \eqs{7} and (\ref{dot})
\begin{equation}
\dot M =
G^2\pi M^2{\varrho_\infty \over a^3_\infty} \left( y-x
\exp \left( -x\gamma \right) \right)^2 \left( \frac{a_\ast^2}{a_\infty^2}
\right)^\frac{(5 - 3 \Gamma)}{2(\Gamma - 1)} .
\label{dotm}
\end{equation}
Now, one can can show in a way similar to that applied in the case of
relativistic accretion \cite{Kinasiewicz, KMM} that under the previously assumed
conditions and $x\gamma \ll 1$ one has
$\varrho_\infty = \chi_\infty \left( M-M_\ast \right) =
M\chi_\infty \left( 1-y\right)$ for $\Gamma \in (1, 5/3 -\delta )$ with some
small $\delta $.

We shall describe briefly main stages of this calculation. It is easy to see,
that \eq{3} yields
\begin{equation}
{a^2(R)\over a^2_\infty } < 1-{\Gamma -1 \over a^2_\infty } \phi (R).
\label{a2}
\end{equation}
Now observe that $|\phi| < GM / R$; therefore
\begin{equation}
{a^2(R)\over a^2_\infty } < 1+\left( \Gamma -1\right) {GM \over Ra^2_\infty }.
\label{a3}
\end{equation}
Returning now to the expression of the mass density given in (\ref{rho}) and
replacing the ratio $a^2(R)/ a^2_\infty $ by the right hand side of (\ref{a3}),
one arrives at $\varrho \le \varrho_\infty \left( 1+\left( \Gamma -1\right)
{GM \over Ra^2_\infty }\right)^{1/(\Gamma - 1)}$.
Applying this bound in $M-M_\ast =\int_V \ud V\, \varrho $ and invoking the
asymptotic conditions (\ref{cond}) one obtains the needed approximation~---
the equality $\varrho_\infty  \approx M\chi_\infty \left( 1-y\right) $.
The proportionality constant $\chi_\infty $ is roughly the inverse of the volume
of the gas outside of the sonic sphere.

Now notice that the total luminosity $L_0=\phi (R_0) \dot M$. Replacing in
$\dot M$ (see \eq{dotm}) the ratio $a^2_\ast / a^2_\infty $ by the right hand
side of (\ref{theorem2}), one obtains
\begin{equation}
L_0=\phi_0 G^2\pi \chi_\infty {M^3 \over a^3_\infty}\left( 1-y\right)
\left( y-x\exp \left( -x\gamma \right) \right)^2\left( {2\over 5 - 3 \Gamma }
\right)^\frac{(5 - 3 \Gamma)}{2(\Gamma - 1)}.
\label{L1}
\end{equation}
It is convenient to rewrite this equation using the relative luminosity
$x=L_0/L_E$:
\begin{equation}
x= \beta \left( 1-y\right) \left( y-x\exp \left( -x\gamma \right) \right)^2.
\label{l1}
\end{equation}
Here $\beta$ is a dimensionless numerical factor depending only on the boundary
data characterizing the system in question,
$\beta = \chi_\infty \phi_0 \alpha G\pi {M^2 \over a^3_\infty}
\left( {2\over 5 - 3 \Gamma } \right)^\frac{(5 - 3 \Gamma)}{2(\Gamma - 1)}$.
We have assumed that the luminosity $x$ is given and our aim is to find
all possible solutions $y(x)$. It appears useful to start from a critical point
and then show the existence of a bifurcation. Let us define
\begin{equation}
F(x,y)\equiv x-\beta \left( 1-y\right)
\left( y-x\exp \left( -x\gamma \right) \right)^2 .
\label{F}
\end{equation}
Let $(a,b)$ be a zero of $F$, that is $F(a,b)=0$. We will say that $(a,b)$ is
critical if $\partial_yF(x,y)|_{a,b}=0$.

One can show that there exist at least two solutions $y(x,\beta )$ for any
parameter $\beta $ ($0\le \gamma <1$) and for the relative luminosity $x$
smaller than $a$. The obtained results are following.

\section*{Theorem}
\begin{enumerate}
\item There exists a unique critical point $x=a$, $y=b$ of $F$, with $a$ and $b$
satisfying bounds $0<a <b<1$ and the relation
$b={2+a\exp \left( -a\gamma \right) \over 3}$.
\label{Theorem-i}

\item For any $0<x<a$ there exist two solutions $y(x){^+_-}$ bifurcating from
$(a,b)$. They are locally approximated by formulae
\begin{equation}
y{^+_-}=b \pm {\sqrt{(a-x)(b+a\exp (-a\gamma )(1-2a\gamma ))}
\over \sqrt{\beta (b-a\exp (-a\gamma )) (1-a\exp (-a\gamma ))}}.
\label{bif}
\end{equation}
\label{Theorem-ii}

\item The relative luminosity $x$ is extremized at the critical point $(a,b)$.
\label{Theorem-iii}
\end{enumerate}

\section*{Proof}
The two conditions $F=0$ and $\partial_yF=0$ yield two equations
\begin{eqnarray}
&&a-\beta (1-b) (b-a\exp (-a\gamma ))^2=0, \nonumber\\
&& b-a\exp (-a\gamma )=2(1-b).
\label{e}
\end{eqnarray}
From \eqs{e} one immediately obtains
\begin{equation}
b={2+a\exp\left( -a\gamma \right) \over 3}.
\label{e0}
\end{equation}
Let us remark that from the above one can get also
\begin{equation}
a=4\beta (1-b)^3;
\label{e00}
\end{equation}
inserting that into the second equation in (\ref{e}), one arrives at
\begin{equation}
b={2\over 3} + {4\beta \over 3}(1-b)^3\exp \left(-4\beta (1-b)^3\gamma \right) .
\label{e1}
\end{equation}
Both sides of this equation are continuous functions of $b$ and at $b=0$ the
right hand side of (\ref{e1}) is bigger than the left hand side, while at $b=1$
the opposite holds true. That guarantees the existence of a solution. Closer
investigation allows one to conclude that the critical solution is unique. This
fact can be used in order to show that solutions $y(x)$ bifurcating from $(a,b)$
extend onto the whole interval $x\in (0, a)$. The argument relies on the use of
the implicit function theorem and the uniqueness of the critical point $(a,b)$.
The specific form of a solution close to a critical point can be obtained in a
standard way. Put $x=a+ \epsilon$, $y=b + y(\epsilon )$ into $F(x,y)=0$ and
expand $F$ keeping a few terms of lowest order. One obtains a reduced algebraic
equation (known in the mathematical literature as the Lyapunov-Schmidt equation)
\begin{eqnarray}
\lefteqn{\left( 1+2\beta \left( 1-b \right) \left( b-a\exp (-a\gamma )\right)
\left( a \exp(-a\gamma )\gamma -\exp (-a\gamma ) \right) \right) \epsilon\times}
\nonumber \\
&&
\beta \left( 3b- 1- 2 a\exp (-a\gamma ) \right) y(\epsilon )^2=0.
\qquad \qquad \qquad \qquad
\label{Lya}
\end{eqnarray}
At the critical point $3b- 1- 2 a\exp (-a\gamma ) = 1- a\exp (-a\gamma )$
(see the second of \eqs{e}) while
\begin{eqnarray}
\lefteqn{2\beta \left( 1-b \right) \left( b-a\exp (-a\gamma )\right)
\left( a\exp (-a\gamma )\gamma -\exp (-a\gamma ) \right) =} \nonumber\\
&&
=2a\left( a\exp (-a\gamma )\gamma -\exp (-a\gamma ) \right)/(b-a\exp (-a\gamma )).
\label{e2}
\end{eqnarray}
Inserting (\ref{e2}) into (\ref{Lya}) and finding $y(\epsilon )$ from the latter
immediately leads to the approximate solution of (\ref{Theorem-ii}).

The proof of the third part of the Theorem goes as follows. Let $(x_0,y_0)$ be
a non-critical solution of $F(x,y)=0$ with the domain belonging to a subset
$xe^{-x}<y$ of the square $0<x<1$, $0<y<1$. The implicit function theorem
guarantees the existence of a curve $x(y)$ such that $F(x(y),y)=0$ for $y$
belonging to some vicinity of $y_0$. Along this curve one has
\begin{equation}
{\ud x\over \ud y}=\beta {\left( y-xe^{-x\gamma }\right)
\left( 3y-2- xe^{-x\gamma } \right) \over
1+ 2\beta (1-y) (y-xe^{-x\gamma }) (e^{-x\gamma }-\gamma xe^{-x\gamma }) }.
\end{equation}
Notice that the denominator is strictly positive while the nominator vanishes
only at critical points. That proves the assertion. Now it is clear from the
form of approximate solutions constructed in part (\ref{Theorem-ii}) that
bifurcating solutions exist only if $x<a$ (our bifurcation is called subcritical
in the mathematical literature); therefore $x=a$ is the extremal value of the
relative luminosity. That ends the proof.

The bifurcating solutions $y(x)$ are shown in the Figure, for the parameter
$\beta =1000$ and under the assumption $\gamma =0$.
\begin{figure}[b]%
\centering%
\includegraphics[height=10.5cm]{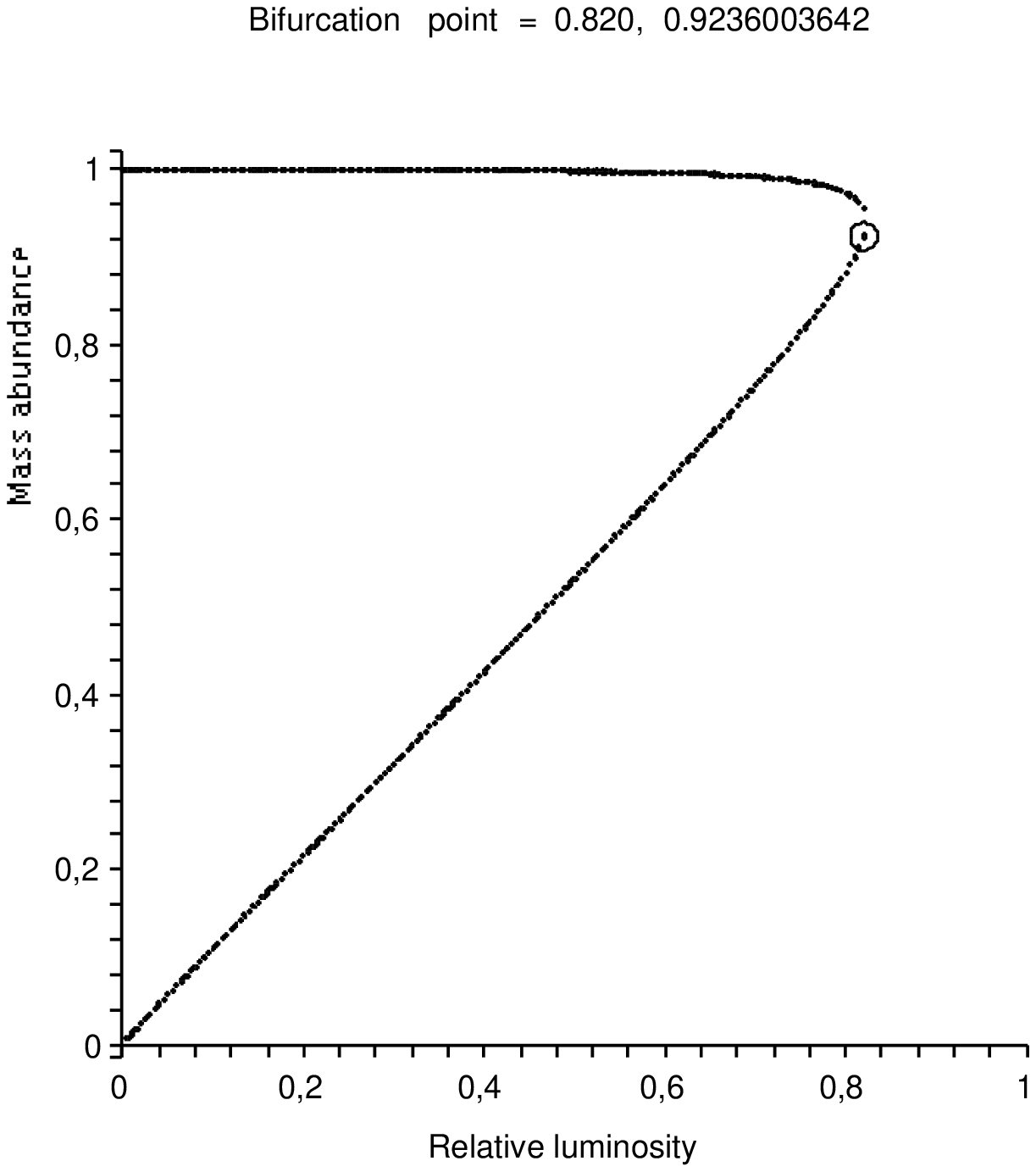}
\caption{The mass abundance is put on the ordinate while the relative
luminosity $x$ is shown on the abscissa. The circle encloses the bifurcation
point $(a,b)$.}
\end{figure}
Notice that coordinates $(a,b)$ of the critical point increase with the increase
of the parameter $\beta $. In order to show that, observe that \eq{e0} implies
${\ud a\over \ud b}=e^{-a\gamma }(1-a\gamma )$. This expression is bigger than
zero, since $a\gamma <1$. Differentiating \eq{e00} with respect to $\beta$ gives
${\ud b\over \ud \beta }=4(1-b)^3/({\ud a\over \ud b} +12 \beta(1-b)^2)>0$;
this inequality is due to the positivity of ${\ud a\over \ud b}$. Now we have
along the critical curve ${\ud a\over \ud \beta }
= {\ud a\over \ud b }{\ud b\over \ud \beta }$; therefore also
${\ud a\over \ud \beta }>0$. Let us remark that if the parameter $\gamma$
vanishes then one can explicitly find $b$ by solving (\ref{e1}). The solution
reads
\begin{equation}
b= {\left( \beta^2+\beta^{3/2} \sqrt{1+\beta }\right)^{2/3}-
\beta \over 2\left( \beta^2+\beta^{3/2} \sqrt{1+\beta }\right) ^{1/3}};
\label{cubic}
\end{equation}
(\ref{e00}) gives then $a(\beta )$ and one can check explicitly that both $a$
and $b$ monotonically increase with the increase of $\beta$.

It is of interest that at critical points the parameter $b$ is not smaller than
$2/3$. From the analysis of the foregoing formulae it is clear that $b$ is
close to the value $2/3$ if $\beta $ is small, \ie, when the relative luminosity
$a$ is small (notice that $a<\beta $). The value $2/3$ appears in the study of
the mass accretion in irradiating systems \cite{Kinasiewicz}. The fluid
abundance is equal to $1-b$; thus we can infer that critical configurations have
less fluid than $1/3$ of the total mass, and that the upper bound $1/3$ is
achieved in the limit of vanishing radiation. The maximal mass accretion rate
does not correspond to $y=2/3$, as in systems with no radiation
\cite{Kinasiewicz}, but to a somewhat larger value.

\section{Conclusions}
In this presentation we have assumed the existence of an accreting system
satisfying particular boundary conditions. We would like to stress out that
(\ref{Theorem-i}) the detailed behaviour of the accreting flow is given by the
integro-differential nonlinear \eqs{1}-(\ref{3}) and (\ref{Theorem-ii}) due to
our boundary conditions some essential features can be deduced from the
algebraic equation (\ref{l1}). The question arises whether this projection of
the original equations onto the algebraic problem in fact takes place.
We checked numerically that there do exist appropriate solutions and that
the relative error made in the above approximations is of the order of
$10^{-3}$. An explicit example can be found in \cite{Karkowski06a}.

It is illuminating to repeat the preceding discussion in the case when
$\gamma \ll 1$. A number of simplification occurs now, since one can approximate
the term $x\exp (-x\gamma )$ $x$. From \eqs{e00} and (\ref{cubic}) one can see
that $a$ can be as close as one wishes to $1$ for sufficiently $\beta$.
Expressing that conclusion in more intuitive terms: the total luminosity of the
accreting system $L_0$ approaches the Eddington luminosity $L_E$ if the
parameter $\beta $ goes to infinity. In particular, from the expression for
$\beta$, if the product $(M/M_0)^2/(a_\infty /a_{0\infty })^3$
(where $M_0$ and $a_{0\infty }$ are some reference quantities) is large enough,
then $L_0$ is close to $L_E$. At the critical point there is only one solution.
The two bifurcation solutions are characterized by the luminosity $x<a$ and
their (central) mass parameters $y_1$ and $ y_2$ can be markedly different
only for $x \ll a$. This can be seen from the point (\ref{Theorem-iii}) of the
Theorem and confirmed by the Figure, which shows the branching of the two
solutions from a critical point in a class of examples. If a given system
radiates with the luminosity close to its critical luminosity (and, in
particular, to the Eddington limit), then the cores corresponding to the
bifurcating solutions have similar masses. The necessary and sufficient
conditions for having two accreting compact stars with hard surfaces that have
masses satisfying condition $M_1 \ll M_2$ is that the luminosity is much smaller
than $L_E$ or the critical luminosity $a$, respectively.

\ack
This paper has been partially supported by the MNII grant 1PO3B 01229.

\end{document}